\documentclass[fleqn,twoside]{article}
\usepackage{espcrc2}

\usepackage{graphicx}
\usepackage[figuresright]{rotating}

\newcommand{\AmS}{{\protect\the\textfont2
  A\kern-.1667em\lower.5ex\hbox{M}\kern-.125emS}}

  \newcommand{\be}{\begin{equation}}\newcommand{\ee}{\end{equation}}
\newcommand{\bea}{\begin{eqnarray} }\newcommand{\eea}{\end{eqnarray}}
\newcommand{\beaa}{\begin{eqnarray}}\newcommand{\eeaa}{\end{eqnarray}}
\newcommand{\ba}{\begin{array}}\newcommand{\ea}{\end{array}}
\newcommand{\bit}{\begin{itemize}}\newcommand{\eit}{\end{itemize}}
\newcommand{\ben}{\begin{enumerate}}\newcommand{\een}{\end{enumerate}}

\def\lf{\left}
\def\non{\nonumber}
\def\pa{\partial}\def\ran{\rangle}
\def\ri{\right}

\def\al{\alpha}\def\ga{\gamma}

\def\te{\theta}
\def\si{\sigma}

\def\lf{\left}

\def\non{\nonumber}
\def\pa{\partial}\def\ran{\rangle}

\def\ri{\right}

\def\al{\alpha}\def\ga{\gamma}

\def\te{\theta}

\def\si{\sigma}

\def\1{{_{1}}}\def\2{{_{2}}}
\def\nof{:\;\!\!\;\!\!:}
\def\wQ{Q}
\def\wwQ{Q}
\title{On flavor violation for massive and mixed neutrinos}

\author{M. Blasone \address[MCSD]{Dipartimento di Matematica e Informatica,
 Universit\`a di Salerno and Istituto Nazionale di Fisica Nucleare,
 Gruppo Collegato di Salerno, 84100 Salerno, Italy,}, A. Capolupo \addressmark,
        C.R. Ji\address{Department of Physics, North Carolina State
University, Raleigh, NC 27695-8202, USA}
        and
        G. Vitiello \addressmark[MCSD]}

\begin{document}

\begin{abstract}
 We discuss flavor charges and states for interacting
mixed  neutrinos in QFT. We show that the  Pontecorvo states
are not eigenstates of the flavor charges. This implies that their
use in describing the flavor neutrinos produces a violation of
lepton charge conservation in the production/detection vertices.
The flavor states defined as eigenstates of the
flavor charges give the correct representation of mixed neutrinos
in charged current weak interaction processes.
\vspace{1pc}
\end{abstract}

\maketitle

In this report we analyze the definition of the flavor charges in
the canonical formalism for interacting (Dirac) neutrinos with mixing.
On this basis, we study the
flavor states for mixed neutrinos in the QFT formalism
\cite{BV95}-\cite{Blasone:2005ae} and in the
Pontecorvo formalism \cite{Pontecorvo:1957cp}-\cite{Beuthe:2001rc}.
We show that Pontecorvo mixed states are not
eigenstates of the neutrino flavor charges and we estimate how much
the leptonic charge is violated on these states.

A realistic description of flavor neutrinos starts by taking into account
the (charged current) weak interaction processes in which they are created,
together with their charged lepton counterparts.
In the Standard Model, flavor is strictly conserved in the
production and detection vertices of
such interactions. The flavor violations are due only to loop corrections and are thus
expected to be extremely small \cite{Casas:2001sr}.
Therefore, we define the flavor neutrino
states as eigenstates of flavor charges. This is obtained in a QFT treatment where
the flavor charges are defined in the usual way from the symmetry
properties of the neutrino Lagrangian.

Here we consider the decay process $ W^{+} \rightarrow e^{+} + \nu_{e}$ and
we study the case where the neutrino mixing is present. We consider for simplicity
the case of two generations. After spontaneous symmetry breaking, the relevant terms of the
Lagrangian density for charged current weak interaction are
$ {\cal L}  =   {\cal L}_{0}  +{\cal L}_{int}\,,$
where  the free lepton Lagrangian ${\cal L}_{0}$
\bea\label{Leptlagrangian2}\non
{\cal L}_{0} & = &
  \lf({\bar \nu_e}, {\bar \nu_\mu}\ri)
\lf( i \ga_\mu \pa^\mu -
M_{\nu} \ri)  \lf(\ba{c}\nu_e\\ \nu_\mu\ea\ri)\,
\\& + &
 \lf({\bar e}, {\bar \mu}\ri)
\lf( i \ga_\mu \pa^\mu -
M_{l} \ri)  \lf(\ba{c}e\\ \mu\ea\ri)\,,
\eea
includes the neutrino non-diagonal mass matrix $M_{\nu}$ and the mass matrix of charged leptons
$M_{l}$:
\bea \non
 M_{\nu} =  \lf(\ba{cc}m_{\nu_e} & m_{\nu_{e\mu}}
\\ m_{\nu_{e\mu}} & m_{\nu_\mu}\ea\ri), \;\;\;
M_l =  \lf(\ba{cc}m_e &0 \\
0 & m_\mu\ea\ri).
\eea
${\cal L}_{int}$ is the interaction Lagrangian given by \cite{PDG}
\bea\label{L-interact}\non
 {\cal L}_{int} & = &  \frac{g}{2\sqrt{2}} \Big [ W_{\mu}^{+}(x)\,
\overline{\nu}_{e}(x)\,\gamma^{\mu}\,(1-\gamma^{5})\,e(x)
\\\non & + &
W_{\mu}^{+}(x)\,
\overline{\nu}_{\mu}(x)\,\gamma^{\mu}\,(1-\gamma^{5})\,\mu(x) + h.c. \Big].
\eea

 ${\cal L}$ is invariant under the global phase transformations:
\bea \label{ephase}  e(x) \rightarrow   e^{i
\alpha}e(x) \,, \qquad \nu_{e}(x) \rightarrow  e^{i \alpha} \nu_{e} (x)\,,
\eea
together with
\bea \label{muphase}
 \mu(x) \rightarrow   e^{i
\alpha}\mu(x) \,, \qquad \nu_{\mu}(x) \rightarrow   e^{i
\alpha} \nu_{\mu} (x)\,.
  \eea
These are generated by the charges $Q_{e}(t)$, $Q_{\nu_{e}} (t)$,
$Q_{\mu}(t)$ and  $Q_{\nu_{\mu}}(t)$ respectively, where
\bea\label{Qflav}
 Q_{\nu_{e}} (t) & = &  \int d^{3}{\bf x} \,
\nu_{e}^{\dag}(x)\nu_{e}(x)\,,
\\  \label{QflavLept}
 Q_{\nu_{\mu}}(t) & = & \int d^{3}{\bf
x} \, \nu_{\mu}^{\dag}(x) \nu_{\mu}(x) \,. \eea
Similar expressions hold for $Q_{e}$, $Q_{\mu}$.
The invariance of the Lagrangian is then expressed by
$ [Q_l^{tot}\,, \, {\cal L} (x)] = 0\,, $
which guarantees the conservation of total lepton number. Here,
$Q_l^{tot}$ is the total Noether (flavor) charge:
\bea Q_{l}^{tot} = Q_{\nu_{e}}(t) + Q_{\nu_{\mu}}(t) + Q_{e}(t) +
Q_{\mu}(t)
\eea
Note that the presence of
the mixed neutrino mass term, i.e.  the non-diagonal mass matrix
$M_\nu$, prevents the invariance of the Lagrangian ${\cal L}_0$
under the separate phase transformations (\ref{ephase}) and
(\ref{muphase}).
Consequently, $Q_{\nu_{e}}$ and $Q_{\nu_{\mu}}$ are time dependent.
However,  family lepton numbers are still good quantum numbers if the neutrino
production/detection vertex can be localized within a region much smaller than the
region where flavor oscillations take place.
This is what happens in practice, since
typically the spatial extension of the neutrino source/detector is much smaller than the
neutrino oscillation length.

We now proceed to define the flavor states as
eigenstates of the flavor charges $Q_{\nu_e}$ and $ Q_{\nu_\mu}$.
Till now, our considerations have been essentially classical.
In order to define the eigenstates of the above charges, we quantize the fields with
definite masses as usual.
Then, the  normal ordered charge operators for free neutrinos $\nu_{1}$,
$\nu_{2}$ are
\bea
:Q_{\nu_{i}}: & \equiv &
\int d^{3}{\bf x} \, : \nu_{i}^{\dag}(x)\;\nu_{i}(x):
\\\non & = & \sum_{r}
\int d^3 {\bf k} \, \lf( \al^{r\dag}_{{\bf k},i}
\al^{r}_{{\bf k},i}\, -\, \beta^{r\dag}_{-{\bf
k},i}\beta^{r}_{-{\bf k},i}\ri),
\eea
where $i=1,2$ and $:..:$ denotes normal ordering with respect to the vacuum $|0\ran_{1,2}$.
The neutrino states with definite masses defined as
\bea
|\nu^{r}_{\;{\bf k},i}\ran =
\al^{r\dag}_{{\bf k},i} |0\ran_{1,2}, \qquad i=1,2,
\eea
are then eigenstates of $Q_{\nu_1}$ and $Q_{\nu_2}$, which can be identified with the
lepton charges of neutrinos in the absence of mixing.

The situation is more delicate when the mixing is present. In such a
case, the flavor neutrino states have to be defined as the
eigenstates of the flavor charges $Q_{\nu_\sigma}(t)$ (at a given
time). The relation between the flavor charges in the presence of
mixing and those in the absence of mixing is given by
\bea\label{carichemix1}
&& {}\hspace{-.5cm}Q_{\nu_e}(t) =  \cos^2\te\;  Q_{\nu_1} + \sin^2\te \; Q_{\nu_2}
\\ \non
&&  {}\hspace{-.5cm}
+ \sin\te\cos\te \int d^3{\bf x} \lf[\nu_1^\dag (x) \nu_2(x) + \nu_2^\dag(x) \nu_1(x)\ri]\,,
\eea
and similarly for $Q_{\nu_\mu}(t)$.
Notice that the presence of the last term in Eq.(\ref{carichemix1})
forbids the construction of eigenstates of
the $Q_{\nu_\sigma}(t)$, $\sigma = e, \mu$, in the Hilbert space
 for free fields ${\cal H}_{1,2}$.

The normal ordered flavor charge operators for mixed neutrinos are then written as
\bea\label{cariche}
&& {}\hspace{-.8cm} \nof \wwQ_{\nu_\sigma}(t) \nof\, \equiv \,\int d^{3}{\bf x}\,
\nof \nu_{\sigma}^{\dag}(x)\;\nu_{\sigma}(x) \nof \
\\\non
&& {}\hspace{-.8cm} =  \sum_{r}
\int d^3 {\bf k} \, \lf( \al^{r\dag}_{{\bf k},\si}(t)
\al^{r}_{{\bf k},\si}(t)- \beta^{r\dag}_{-{\bf
k},\si}(t) \beta^{r}_{-{\bf k},\si}(t)\ri),
\eea
where $\sigma=e,\mu$, and $\nof ... \nof\,$
denotes normal ordering with respect to the flavor vacuum $|0\ran_{e,\mu}$,
the vacuum for the Hilbert spaces of interacting fields ${\cal H}_{e,\mu}$.
 At finite volume it is given by: $ |0(t) \rangle_{e,\mu} = G^{-1}_{\bf \te}(t)\;
|0 \rangle_{1,2}$, where $G_{\bf \te}(t)$ is the mixing generator.
Indeed, the mixing transformations can be written as
$ \nu_{\sigma}^{\alpha}(x) = G^{-1}_{\bf \te}(t)\;
\nu_{i}^{\alpha}(x)\; G_{\bf \te}(t)$,
where $(\sigma,i)=(e,1) , (\mu,2)$.
In a similar way, flavor annihilators, relative
to the fields $\nu_{\sigma}(x)$  at each time are
expressed as:
$
\alpha _{{\bf k},\sigma}^{r}(t) = G^{-1}_{\bf
\te}(t)\;\alpha _{{\bf k},i}^{r}(t)\;G_{\bf \te}(t)$,
$\beta _{{\bf k},\sigma}^{r}(t) = G^{-1}_{\bf \te}(t)\;\beta
_{{\bf k},i}^{r}(t)\;G_{\bf \te}(t)$,
where $\alpha ^{r}_{{\bf k},i}$ and $ \beta ^{r }_{{\bf
k},i}$, $ i=1,2 \;, \;r=1,2$ are the annihilation operators for  $|0\rangle_{1,2}$.

Eq.(\ref{cariche}) shows that the flavor charges are diagonal
 in the flavor annihilation/creation operators.
Note that
$
 \nof \wwQ_{\nu_\sigma}(t) \nof \; = \,G_\theta^{-1}(t)  :\wQ_{\nu_j} :
G_\theta(t),
$
 with $(\sigma,j) = (e,1),(\mu,2),$ and
\bea \non
\nof {Q}_{\nu} \nof & = &
 \nof {\wwQ}_{\nu_e}(t) \nof \;+\; \nof{\wwQ}_{\nu_\mu}(t)\nof
\\
& =& :{\wQ}_{\nu_{1}}:\; +\; :\wQ_{\nu_{2}}: \; = \; :\wQ_{\nu}:\,.
\eea
The flavor states are defined as eigenstates of the flavor charges
$\wwQ_{\nu_\sigma} $ at a reference  time $t=0$:
\bea\label{flavstate}
 |\nu^{r}_{\;{\bf k},\si}\ran \equiv
\al^{r\dag}_{{\bf k},{\sigma}}(0) |0(0)\ran_{{e,\mu}}, \qquad \si =
e,\mu .
\eea

Let us turn now to the  Pontecorvo  states \cite{Pontecorvo:1957cp}-\cite{Beuthe:2001rc}
\begin{eqnarray} \label{nue0a}
|\nu^{r}_{\;{\bf k},e}\rangle_P &=& \cos\theta\;|\nu^{r}_{\;{\bf k},1}\rangle \;+\;
\sin\theta\; |\nu^{r}_{\;{\bf k},2}\rangle \,,
\\ [2mm] \label{nue0b}
|\nu^{r}_{\;{\bf k},\mu}\rangle_P &=& -\sin\theta\;|\nu^{r}_{\;{\bf k},1}\rangle \;+\;
\cos\theta\; |\nu^{r}_{\;{\bf k},2}\rangle \,.
\end{eqnarray}
They are clearly {\em not} eigenstates of the flavor charges \cite{Blasone:2005ae}
as can be seen from Eqs.(\ref{carichemix1}) and (\ref{cariche}).
In order to estimate how much the lepton charge is violated in the
usual quantum mechanical states, we consider the expectation values
of the flavor charges on the Pontecorvo states. We obtain, for the
electron neutrino charge,
\bea\label{Qnu}\non
 \;_{P}\langle\nu^{r}_{\,{\bf k},e}| \nof \wwQ_{\nu_e}(0)\nof
|\nu^{r}_{\,{\bf k},e}\rangle_{P} = \cos ^{4}\theta +
\sin^{4}\theta
\\  +\,  2 |U_{\bf k}| \sin^{2}\theta \cos^{2}\theta
+\sum_{r}\int d^{3}{\bf k}\,, \eea
where $  |U_{{\bf k}}|\equiv u^{r\dag}_{{\bf k},1}
u^{r}_{{\bf k},2} = v^{r\dag}_{-{\bf k},1} v^{r}_{-{\bf k},2} $ and
 \bea\label{Q0}
\;_{1,2}\langle 0 |\nof \wwQ_{\nu_e}(0) \nof| 0 \rangle_{1,2} = \sum_{r}\int d^{3}{\bf k}\,.
\eea
 The infinities in Eqs.(\ref{Qnu}) and (\ref{Q0}) may be removed by considering
 the expectation values of  $:\wQ_{\nu_\sigma}(t):$, i.e. the
normal ordered  flavor charges with respect to the mass vacuum
$|0\ran_{1,2}$. Then,
\bea \;_{1,2}\langle 0 |: \wQ_{\nu_e}(0): | 0 \rangle_{1,2} = 0\,.
\eea
However,  \cite{Blasone:2005ae}
\bea\label{caric}
\non \;_{P}\langle\nu^{r}_{\,{\bf k},e}|:
\wQ_{\nu_e}(0): |\nu^{r}_{\,{\bf k},e}\rangle_{P}  =  \cos ^{4}\theta
+ \sin^{4}\theta
\\
 + 2 |U_{\bf k}| \sin^{2}\theta \cos^{2}\theta < 1,
\eea
\bea
\label{A1}\non
 \;{}_{P}\langle\nu^{r}_{\,{\bf k},e}|:
\wQ_{\nu_\mu}(0): |\nu^{r}_{\,{\bf k},e}\rangle_{P} = 2
\sin^{2}\theta \cos^{2}\theta \times
\\
\times (1 - |U_{\bf k}|) > 0,
\eea
for any $ \theta \neq 0,\;  m_{1} \neq m_{2},\; {\bf k} \neq
0.$

In conclusion, the correct flavor states describing the neutrino oscillations must be those
defined in Eq.(\ref{flavstate}).

We also note that similar results about flavor violation for massive neutrinos
have been recently discussed in Ref.\cite{Nishi:2008sc}, although from a
different point of view.


\end{document}